\documentclass[12pt]{iopart}
\usepackage{iopams}  
\usepackage{graphicx}  
\usepackage{amsfonts}  
\usepackage{mathrsfs}  

\newcommand{\Hb}{{\hat{\mathcal H}}_{\beta}}

\newcommand{\ep}{\epsilon} 
\newcommand{\ps}{\ps}

\newcommand{\Up}{{\hat{\mathcal U}}}

\newcommand{\Np}{{\hat{\mathcal N}}}

\newcommand{\Ub}{\Up_{\beta}}

\newcommand{\sign}[1]{\rm{sign}(#1)} 
\newcommand{\mean}[1]{\left\langle#1\right\rangle} 
\newcommand{\abs}[1]{\left|#1\right|} 
\newcommand{\op}[1]{\hat{#1}} 
  
\begin{document} 
 
\title[The role of quasi-momentum in the resonant dynamics of the AOKR]
{The role of quasi-momentum in the resonant dynamics of 
the atom--optics kicked rotor} 
 
\author{Sandro Wimberger \dag\footnote[3]
{Corresponding author's e-mail: saw@pks.mpg.de}  
and Mark Sadgrove\ddag} 
 
\address{\dag CNR-INFM and Dipartimento di Fisica Enrico Fermi,  
Universit\`{a} di Pisa, Largo Pontecorvo 3, 56127 Pisa, Italy} 
 
\address{\ddag Department of Physics,   
University of Auckland, Private Bag 92019, Auckland, New Zealand}

\begin{abstract} 
We examine the effect of the initial atomic momentum distribution on the 
dynamics of the atom--optical realisation of the quantum kicked rotor.  
The atoms are kicked by a pulsed optical lattice, the periodicity of which 
implies that quasi--momentum is conserved 
in the transport problem. We study and compare 
experimentally and theoretically two resonant limits  
of the kicked rotor: in the vicinity of the quantum resonances and in the 
semiclassical limit of vanishing kicking period. It is  
found that for the \textit{same} experimental 
distribution of quasi--momenta, significant deviations  
from the kicked rotor model are induced 
close to quantum resonance, while close to the  
classical resonance (i.e. for  
small kicking period) the effect of the quasimomentum vanishes. 
 
\end{abstract} 
 
 
 
 
\section{Introduction} 
The past decade has brought fascinating advances in the  
preparation and control of single particles \cite{oneatom}.  
Atoms can now be cooled down to a level where the effect  
of a single photon recoil can be measured experimentally  
\cite{Monroe1990}. Single atom dynamics can thus 
be controlled with high precision by introducing an external  
field in the form of  
an optical potential \cite{Moore1995,BEClattice}. 
 
A particular example of such a system, the atom--optics  
kicked rotor, has shed light 
on interesting and paradigmatic quantum effects including  
dynamical localisation \cite{Moore1995} 
and quantum resonance \cite{Oskay2000,darcy1,darcy2,Sadgrove2004}.  
In all such experiments, control of the initial conditions  
in phase space is essential. 
In particular, the impact of different momentum classes  
on the dynamics near quantum resonance was explained recently  
\cite{darcy3,Wimberger2003}. The atoms are kicked 
by a spatially periodic potential which is pulsed  
on at a certain frequency. As dictated 
by standard Bloch theory, the spatial periodicity  
implies that the quasi--momentum for the centre-of-mass motion of each atom 
is conserved during the evolution. Quasi--momentum  
is an intrinsically quantum variable which 
arises due to the translational symmetry of the problem \cite{Ashcroft}.  
Since experiments with cold atoms  
typically use a broad, continuous distribution of  
quasi--momenta, the experimental data  
represents a result averaged over this initial distribution  
\cite{darcy3,Wimberger2003,Sadgrove2005,Raizen1999}. 
 
The averaging over different momentum classes leads to  
significant deviations from the  
standard $\delta$--kicked rotor model \cite{rotor,Izr} which 
typically does not consider the additional control  
parameter introduced by the quasi--momentum. 
Such deviations have been experimentally observed,  
in particular at quantum resonance
\cite{darcy3} and have been explained theoretically  
by means of a new pseudo--classical  
model introduced in \cite{FGR} and applied to the  
usual $\delta$--kicked rotor in  
\cite{Wimberger2003,Wimberger2004}. 
 
In this paper, we use the same theoretical formalism to  
expose the innate similarities and surprising  
differences between the limit in which the exact quantum  
resonant driving is approached and the limit 
of vanishing kicking period. The former limit can be  
described using the pseudo--classical model from 
\cite{Wimberger2003,Wimberger2004} 
(with an effective Planck constant defined by the detuning  
from exact resonance), whilst the latter limit 
is the usual classical limit of the kicked rotor  
(with the scaled kicking period as the effective 
Planck constant). Our theoretical analysis of the  
experimental data focuses on the role of the quasi--momentum, 
which proves to be quite different in the two  
``classical'' limits studied here. 
 
\section{The Atom--Optics Kicked Rotor} 
 
We consider a system of Caesium atoms in an optical  
standing wave (with wave number $k_L$)  
which is $\delta$--pulsed  
with period $\tau$. For sufficiently large detuning  
from the atomic absorption line, the 
Hamiltonian for an atom is given by \cite{Graham1992} 
\begin{equation}  
H(t') =\frac{p^2}{2} + k\cos(z)\sum_{t=0}^{N}  
 \delta (t'-t\tau)\;,  
\label{eq:ham}  
\end{equation}  
where $p$ is the atomic momentum in units of $2\hbar k_L$   
(i.e., in units of two-photon recoils),  
$z$ is the atomic position in units of   
$2k_L$, $t'$ is time and $t$ is the kick number.  
The scaled kicking period $\tau$ is defined by the equation 
$\tau=8E_R T/\hbar$, where $E_R=\hbar^2 k_L^2/2M$ is  
the recoil energy (associated with  
the energy change of a Caesium atom of mass $M$ 
after emission of a photon of wavelength   
$\lambda_L = 2\pi/k_L = 852\;\rm$nm). The kicking  
strength of the system 
is given by $k=V_0\tau/\hbar$ where $V_0$ is the maximum potential depth
created by the optical standing wave \cite{Moore1995,Graham1992}. 
 
Experimentally, momentum kicks are delivered to the atoms  
by an optical lattice which is created by   
a 150mW diode laser injection locked to a lower power  
feedback stabilised 
source at 852 nm. Kicking laser powers of up to  
$30\; \rm mW$ were employed for detunings of $500 \;\rm MHz$ from 
the  $6S_{1/2}(F=4) \rightarrow 6P_{3/2}(F'=5)$  
transition of Caesium. For the experimental results  
presented in this paper, 
the average energy of the atomic ensemble was measured after up to 20 kicks. 
To control the pulse timing, a custom built  
programmable pulse generator was employed 
to gate an acousto--optic modulator which controlled  
the amount of kicking light  
reaching the atomic sample. Timing of the experiment  
was controlled by a real-time, software based 
computer system with a latency on the order of 10$\mu$s.   
 
For the classical resonance experiments reported here,  
the kicking pulse width 
was 320\;ns, whilst for the quantum resonance results,  
a 480\;ns pulse width was used. In 
the classical limit of vanishing kicking period, the  
$\delta$--kick approximation is violated 
in the experiment (although for 
the small kick numbers and kicking strengths used  
here, our results do not  
show deviations from the $\delta$--kick theory  
\cite{Sadgrove2005a,Blumel1986,Raizen1998}). 
As a consequence, it is possible to probe the  
dynamics at exact \emph{quantum} resonance, 
but not at the exact classical limit, since  
the pulse period $\tau$ should always exceed 
the pulse width to ensure a reliable approximation  
to $\delta$--pulses. 
 
The experimental sequence ran as follows:  
Atoms were released from the   
magneto--optical trap~\cite{Monroe1990} and  
then kicked by a series of light pulses. A  
free expansion time of 
12 ms was then allowed followed by ``freezing''  
of the atomic motion 
in optical molasses and subsequent CCD imaging  
of the resultant atomic cloud~\cite{Sadgrove2004}. 
Mean energies are extracted from the raw data  
by calculating the second moment of the experimentally measured  
momentum distribution of the atoms' centre-of-mass motion.  
 
By exploiting the spatial periodicity of the  
Hamiltonian (\ref{eq:ham}), the atomic 
dynamics along the $z$ axis can be reduced to  
that of a rotor on a circle by Bloch's Theorem 
\cite{Wimberger2003}. This introduces the  
additional parameter $\beta \in [0,1)$ which represents 
the atomic quasi--momentum -- a constant of the  
motion by Bloch's theorem. 
The fractional part of the physical momentum $p$ in the units  
given above corresponds to the quasi--momentum 
which is practically uniformly distributed  
in the fundamental Brillouin zone defined by the periodic kick 
potential \cite{Wimberger2003}. 
The one-kick propagation operator for a given  
atom is ~\cite{Wimberger2003} 
\begin{equation}  
\Ub \;=\;e^{-{\rm i} k\cos({\hat \theta})} 
\;e^{-{\rm i}\tau(\Np+\beta)^2/2},  
\label{eq:onecycle}  
\end{equation}  
where $\theta=x$mod$(2\pi)$, and $\Np=-{\rm i} 
d/d\theta$ is the angular momentum operator 
with periodic boundary conditions. 
 
\section{Unifying Classical Description of  
Quantum and Classical Resonance} 
\label{drei} 
 
\begin{figure} 
\centering 
\includegraphics[height=8cm]{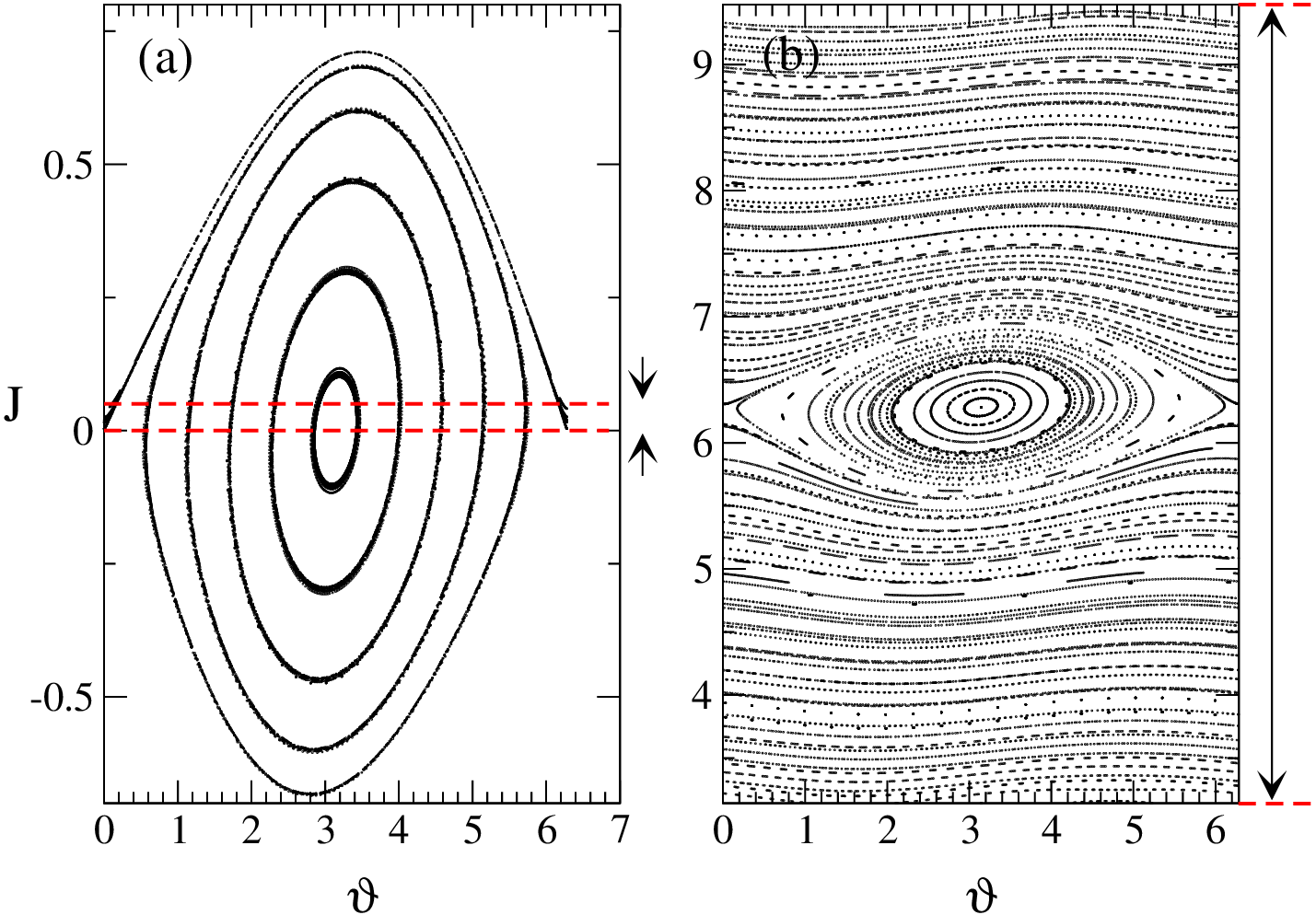} 
\caption{ 
Phase space portrait generated by the  
map~(\ref{eq:epsmap}) for $k=2.5$ and $\epsilon=0.05$. 
The initial angles $\theta_0$ were uniformly  
distributed in $[0,2\pi)$ whilst the 
initial momenta $J_0$ were taken from uniform  
distributions on the two different intervals 
$[0,\epsilon)$ (a) and $[\pi, 3\pi+\epsilon)$  
(b) as shown by the arrows in both figures.
Note that the phase space is $2\pi$-periodic
along the $J$ axis. 
}  
\label{fig:1}. 
\end{figure} 
 
The quantum dynamics in the two semiclassical limits  
studied here is approximated by the following map  
\cite{Sadgrove2005,Wimberger2004}: 
\begin{eqnarray}  
I_{t+1} &=& I_{t} + {\tilde k} \sin (\theta_{t+1}) \;, \nonumber \\ 
\theta_{t+1} &=& \theta_t \pm I_t + \ell\pi + \tau \beta  
\;\; {\mbox{\rm mod}}(2\pi ), 
\label{eq:clmap}  
\end{eqnarray} 
where $\tau=2\pi\ell + \epsilon$ and ${\tilde k}=k|\epsilon|$, 
and $\ell=0,1,2$ ($\pm$ is the sign of $\epsilon$,
and for $\ell =0$ only $+$ is allowed). 
The above map is similar to the well-studied  
Standard Map \cite{Chirikov1979} augmented by the  
term $\tau\beta$ which accounts for the experimental  
quasi--momentum distribution. 
Changing variables to $J= \pm I + \ell\pi + \tau\beta$,  
$\vartheta=\theta+\pi(1-\sign{\epsilon})/2$ 
formally gives the true Standard Map 
\begin{eqnarray} 
J_{t+1} & = & J_t+{\tilde k}\sin (\vartheta_{t+1}) \;, \nonumber \\
\vartheta_{t+1} & = & \vartheta_t+J_t \;.
\label{eq:epsmap} 
\end{eqnarray} 
The mean energy is calculated using the formula  
\begin{equation} 
\mean{E_{t,\epsilon}} = \epsilon^{-2}\mean{I_t^2}/2  
= \epsilon^{-2}\mean{\delta J_t^2}/2,\;\;\;\; \delta J_t = J_t-J_0. 
\end{equation} 
Although the map (\ref{eq:epsmap}) is not explicitly dependent on the  
additional $\beta$ dependent term, 
we note that the initial conditions in momentum space are given by 
$J_0 = \pm I_0 + \pi\ell + \tau\beta$, i.e., they are defined 
by the initial choice of quasi--momentum $\beta$.  
\begin{figure}[htb] 
\centering 
\vspace{1.0cm}
\includegraphics[height=8cm]{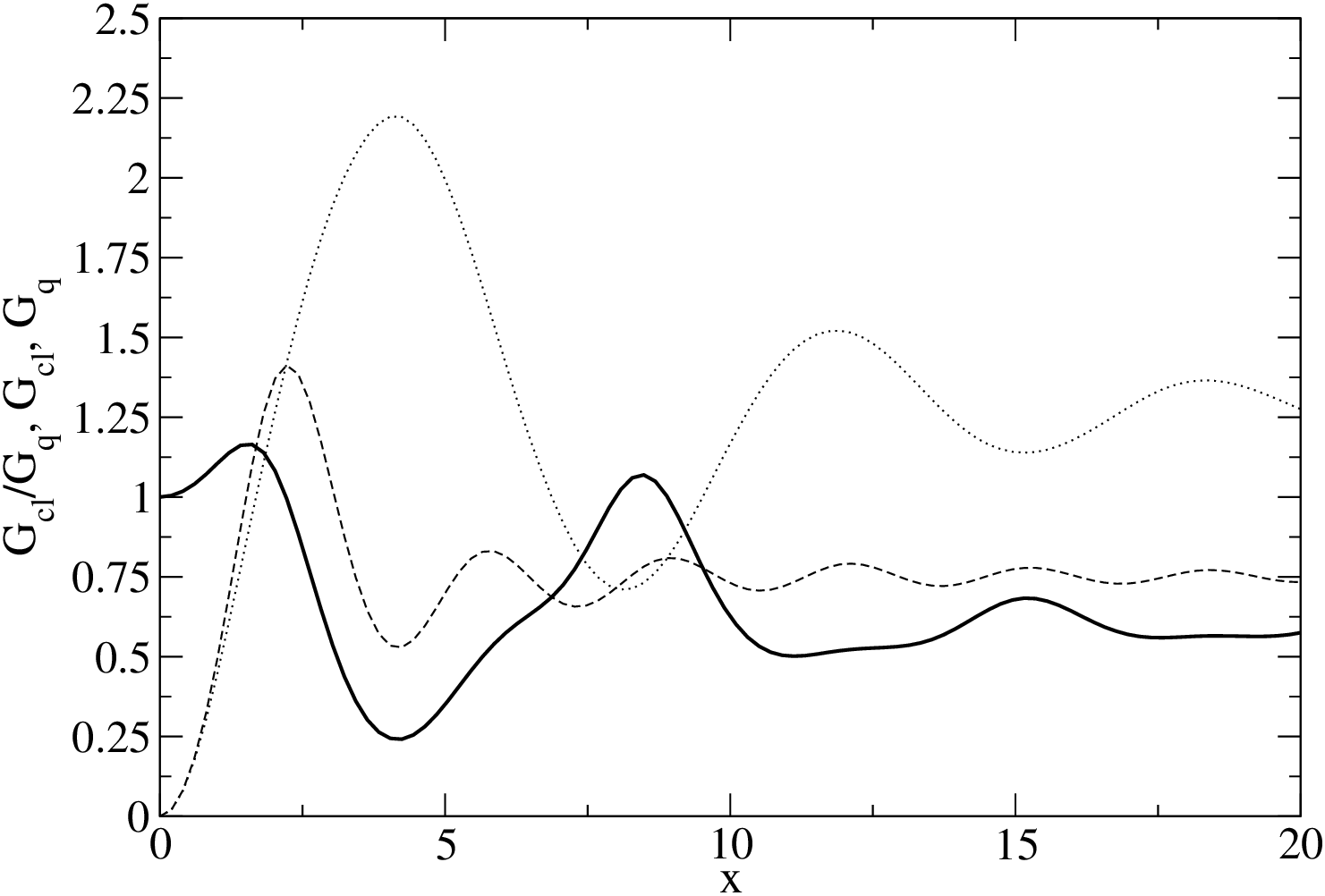} 
\caption{ 
The ratio $G_{\rm cl}/G_{\rm q}$ (solid line) is shown along with the functions 
$G_{\rm cl}$ (dashed line) and $G_{\rm q}$ (dotted line) themselves. The ratio saturates
to a constant for large $x$ after initial oscillations, as the classical and quantum 
resonance peaks decay at the same rate. The differences between the two scaling functions
arise due to the different initial conditions in phase space in the classical and
$\epsilon$--classical limits (see Fig. \ref{fig:1}).}  
\label{fig:2}. 
\end{figure} 

Two a priori  
quite different regimes are described by either of the two maps 
(\ref{eq:clmap}) or (\ref{eq:epsmap}):  
firstly that for $\ell=0$, and 
$\tau\rightarrow 0$ and secondly that for  
$\ell>0$, $\tau\rightarrow 2\pi\ell$, for integer $\ell$. 
In the case where $\ell=0$ we have  
$\epsilon=\tau$ and $J=\tau p$, with the physical momentum 
$p$ in units of two--photon recoils \cite{Sadgrove2005}. 
For integer $\ell > 0$, the map in (\ref{eq:epsmap})  
approximates the dynamics near 
the fundamental quantum resonances occurring  
at $\tau=2\pi\ell$. As shown in Refs.  
\cite{Wimberger2003,Wimberger2004}, the one-kick
propagator (\ref{eq:onecycle}) may be rewritten in the form  
\begin{equation}  
\Ub (t)\;=\;e^{-{\rm i}\tilde k\cos({\hat \theta})/|\ep| } 
\;e^{- {\rm i}\Hb/|\ep| }\;, 
\label{eq:cycleeps} 
\end{equation}  
where $\epsilon=\tau-2\pi\ell$, $\tilde{k}= 
\abs{\epsilon}k$, $\hat{I}=\abs{\epsilon}\Np$ and 
$\op{H}_\beta = \frac{1}{2}\sign{\epsilon}\op{I}^2  
+ \op{I}(\pi\ell + \tau\beta)$. 
Considering $|\epsilon|$ to be an effective  
Planck constant, we see that the 
map given in Eq.~(\ref{eq:clmap}) approximates  
the dynamics induced by (\ref{eq:onecycle}) in both classical 
limits for  $\epsilon\rightarrow 0$. 
 
Figure \ref{fig:1} demonstrates the essential difference between  
the two semiclassical limits studied here. In the case where $\ell=0$  
(see Fig.~\ref{fig:1} (a)), 
 a uniform quasi--momentum distribution on $[0,1)$ leads 
to the initial momenta $J_0$ being  uniformly distributed on 
the interval $[0,\sigma_p\epsilon)$, where $\sigma_p$ is the
characteristic width of the initial atomic momentum distribution in units
of two-photon recoils. Therefore, for $\sigma_p \sim 1$,  
the initial momenta lie entirely within the region of phase  
space dominated by the nonlinear resonance 
island of the Standard Map .  
For $\ell=1$ (see Fig.~\ref{fig:1} (b)), and the {\em same} uniform  
quasi--momentum distribution, the initial momenta  
populate the full unit cell $[\pi,3\pi)$ in the periodic  
phase space which encompasses not only the nonlinear resonance island at 
$J=2\pi$, but also regular ``rotation'' motion beyond it.  
Therefore the same experimental quasi--momentum  
distribution leads to different 
behaviour of the atomic ensemble in the two  
limits of $\ell=0$ and $\ell\neq 0$. 
 
Based on the maps~(\ref{eq:clmap}) and (\ref{eq:epsmap}),  
useful results were previously 
derived for the analysis of experimental data  
\cite{Sadgrove2005,Wimberger2004}. 
These results may be summarised by the following  
single parameter scaling functions which  
differ for the two limits of interest here. 
For $\ell=0$, the scaling function of the  
mean energy close to $\epsilon = \tau=0$ is given by 
\begin{equation}  
\label{eq:scalcl}  
\frac{\langle E_{t,\tau}\rangle}{\langle E_{t,0}\rangle}   
\approx R_{cl}(x) \equiv  \frac{2}{x^2} G_{\rm cl}(x)\;,  
\end{equation}  
with $x=t\sqrt{k\abs{\epsilon}}$  
and the function $G_{\rm cl}$ defined by
\begin{equation}
\label{eq:gcl} 
G_{cl}(x)  \approx \frac{1}{2\pi}\int_0^{2\pi}{\rm d}\theta_0 \overline{J} 
(x,\theta_0, J_0=0)^2\;, 
\end{equation} 
where $\overline{J}\equiv J/\sqrt{{\tilde k}}$ is the momentum of the pendulum 
approximation to the dynamics generated by the map of Eq. (\ref{eq:clmap}) as defined  
previously in Ref. \cite{Wimberger2003,Wimberger2004}. 
 
For $\ell>0$, we have instead close to $\epsilon = 0$
\begin{equation}  
\label{eq:scal}  
\frac{\langle E_{t,\ep}\rangle}  
{\langle E_{t,0}\rangle} \approx R_q(x)\equiv  
1-\Phi_0(x)+\frac{4}{\pi x}G_{\rm q}(x)\;,  
\end{equation}  
with different functions $\Phi_0$ and $G_q$. In this case,
we have
\begin{equation}
G_q(x) \approx 
\frac{1}{8\pi} 
\int_0^{2\pi}{\rm d}\theta_0 \int_{-2}^{2}{\rm d}J_0 
\overline{J}(x,\theta_0, J_0)^2\;. 
\end{equation}
The difference between the two scaling functions $G_{\rm cl}$
and $G_{\rm q}$ may be seen in Fig. \ref{fig:2} where the ratio
of the two functions is plotted along with the functions themselves. 
Although the functions have the same slope for small $x$, their 
forms differ in general and for large $x$, the ratio saturates 
to a constant less than 1. The
difference in the saturation values of the two $G$ functions arises 
from the different initial conditions in the 
the phase space of map \ref{eq:clmap} which apply in the classical
and $\epsilon$--classical limits.
 
In the following section, we compare experimental  
data for the two different cases $\ell=0$ and $\ell=1,2$ 
guided by the theoretical results reviewed in the present section.

\section{Experimental vs. Theoretical Results} 
 
\begin{figure} 
\centering 
\includegraphics[height=8cm]{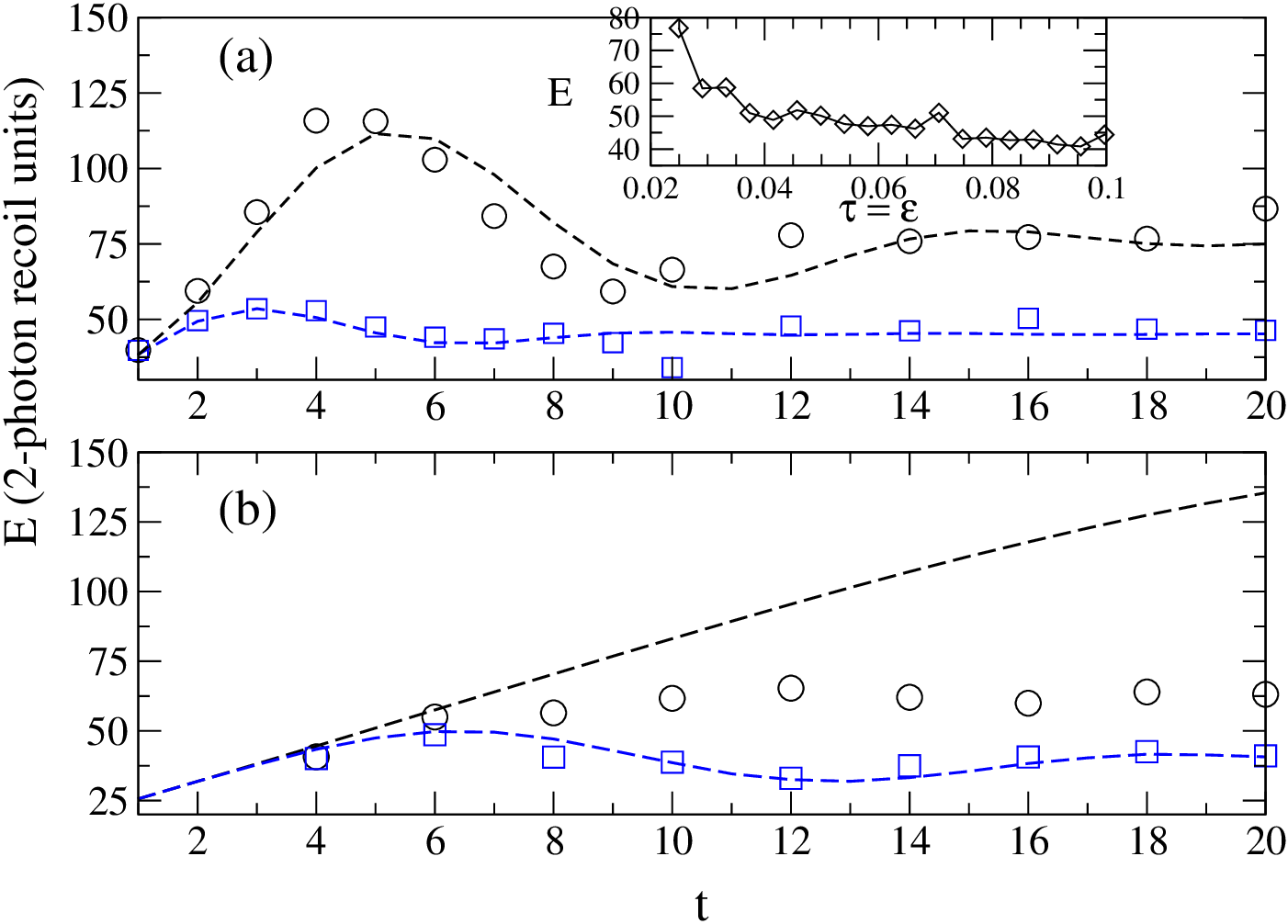} 
\caption{ 
Experimental measurements of the mean energy as a  
function of kick number for $k \approx 5$,  
taken for small values of the detuning $\epsilon$  
in the limits $\tau \to 0$ (a) and  
$\epsilon \to 0$ for $\ell = 1$ (b). In particular, 
we have (a) $\epsilon=0.033$ (circles) and $\epsilon \approx  
0.08$ (squares), and (b) 
$|\epsilon| \lesssim 0.005$ (circles)  
and $\epsilon \approx 0.08$ (squares),   
along with classical simulations using the map (\ref{eq:epsmap})  
(dashed lines). 
We note the oscillatory nature of the energy curve for finite detuning 
$\epsilon$, which may be viewed as a consequence of the  dynamics 
represented by the phase spaces in Fig. \ref{fig:1}. 
The inset in (a) shows a detailed experimental scan of the  
classical resonance peak as $\tau=\epsilon \rightarrow 0$, 
for $k\approx 2.5$ and after $t=5$ kicks. 
}\label{fig:3} 
\end{figure}

In Figure \ref{fig:3}, experimentally measured  
energies close to the classical and quantum resonances 
are plotted against the kick number. In both  
plots of this figure, the observed oscillatory behaviour  
may be understood in terms of the pendulum approximation  
to the dynamics of the map (\ref{eq:epsmap}) as embodied by the 
functions $G_{\rm cl}(x)$ or $G_{\rm q}(x)$.  
\cite{Wimberger2003,Sadgrove2005,Wimberger2004,pra2005}.  
For small times ($t<5$ for the data in Fig. \ref{fig:3} (a)),  
the energy growth near the classical resonance  
is \emph{ballistic}, i.e., the energy grows 
quadratically in time. 

We note that ballistic motion is also predicted to 
occur at quantum resonance for an atomic ensemble 
with a very narrow initial momentum distribution \cite{Duffy2004,pisa}.  
But the broad initial momentum distribution present in cold atom experiments 
as discussed here, typically leads to a uniform distribution of all possible 
values of quasi--momentum \cite{darcy3,Wimberger2003}.  
In terms of the classical model reviewed in the previous section, these 
experimental initial conditions correspond to  
initial momenta distributed over the full phase space cell,  
as shown in Fig. \ref{fig:1} (b).  
The majority of the atoms obey rotational motion 
with almost constant energies (see Fig. \ref{fig:1} (b)),  
whilst only a small sub-class follows 
the motion inside the nonlinear resonance island,  
which for a finite time (depending on the detuning $\epsilon$) supports
ballistic energy growth \cite{Wimberger2003,Wimberger2004}. 
 
The connection between the dynamics in the  
classical limit and that for a quantum particle  
starting from a momentum eigenstate 
is found in the term $\tau\beta$ in the map  
(\ref{eq:clmap}). We see that this term may become zero in either 
of the following limits: $\tau\rightarrow 0$  
or $\beta\rightarrow 0$. In both cases, 
the effect is to regain ballistic energy growth. 
The inset in Fig. \ref{fig:3} (a) 
shows a detailed scan of the mean energy near  
the classical resonance as $\tau\rightarrow 0$ which 
emphasises the rapid energy growth seen in this  
regime associated with the ballistic classical resonance. 
 
Figure \ref{fig:3} (b) shows  
mean energy measurements at exact quantum resonance (circles) 
and for $\epsilon \approx 0.08$ along with  
$\epsilon$--classical simulation results (dashed lines). 
For {\em the same experimental momentum distribution},  
only linear mean energy growth is predicted to occur 
at exact quantum resonance. Additionally, the data 
shown here demonstrate a practical problem which  
arises from the uniform distribution of quasi--momenta 
over the first Brillouin zone. Because only  
the quasi--momentum classes $\beta \approx 1/2$ (for $\ell =1$) and
$\beta \approx 0,1/2$ (for $\ell =2$) experience quantum resonant dynamics 
\cite{darcy3,Wimberger2003,Izr},  
only a small number of resonant atoms are responsible for the linear growth 
of the ensemble mean energy. The measurement of the mean energy at exact 
quantum resonance is therefore experimentally very challenging 
since the signal-to-noise ratio is low for  
the small population of resonant atoms 
\cite{Oskay2000,darcy2,darcy3,Wimberger2003}.  
This is the most likely cause of the apparent  
saturation of energy growth in the quantum resonance 
case as seen in Fig. \ref{fig:3} (b) where the experimental mean energy (circles)
noticeably deviates from the expected linear growth (dashed line).
Indeed, inspection of the experimental momentum distributions for the on--resonance data   
reveals that the characteristic ballistic wings associated
with resonant atoms \cite{darcy3} are not resolved for kick numbers greater than about $6$ in these
experiments. 

By comparison with the data in \ref{fig:3} (a) for the classical resonance, 
we see that, even though the maximum energy 
is much larger than that measured at quantum resonance for the same number of kicks,
 the initial quadratic mean energy growth can easily be 
resolved since practically 
the {\em entire atomic ensemble} experiences  
resonant energy growth in this regime. This is precisely because 
as $\tau$ tends to zero, the $\beta$  
dependence of the map (\ref{eq:clmap}) is removed as the term 
$\tau\beta$ vanishes at the same rate as $\tau$. 
 
\begin{figure} 
\centering 
\includegraphics[height=8cm]{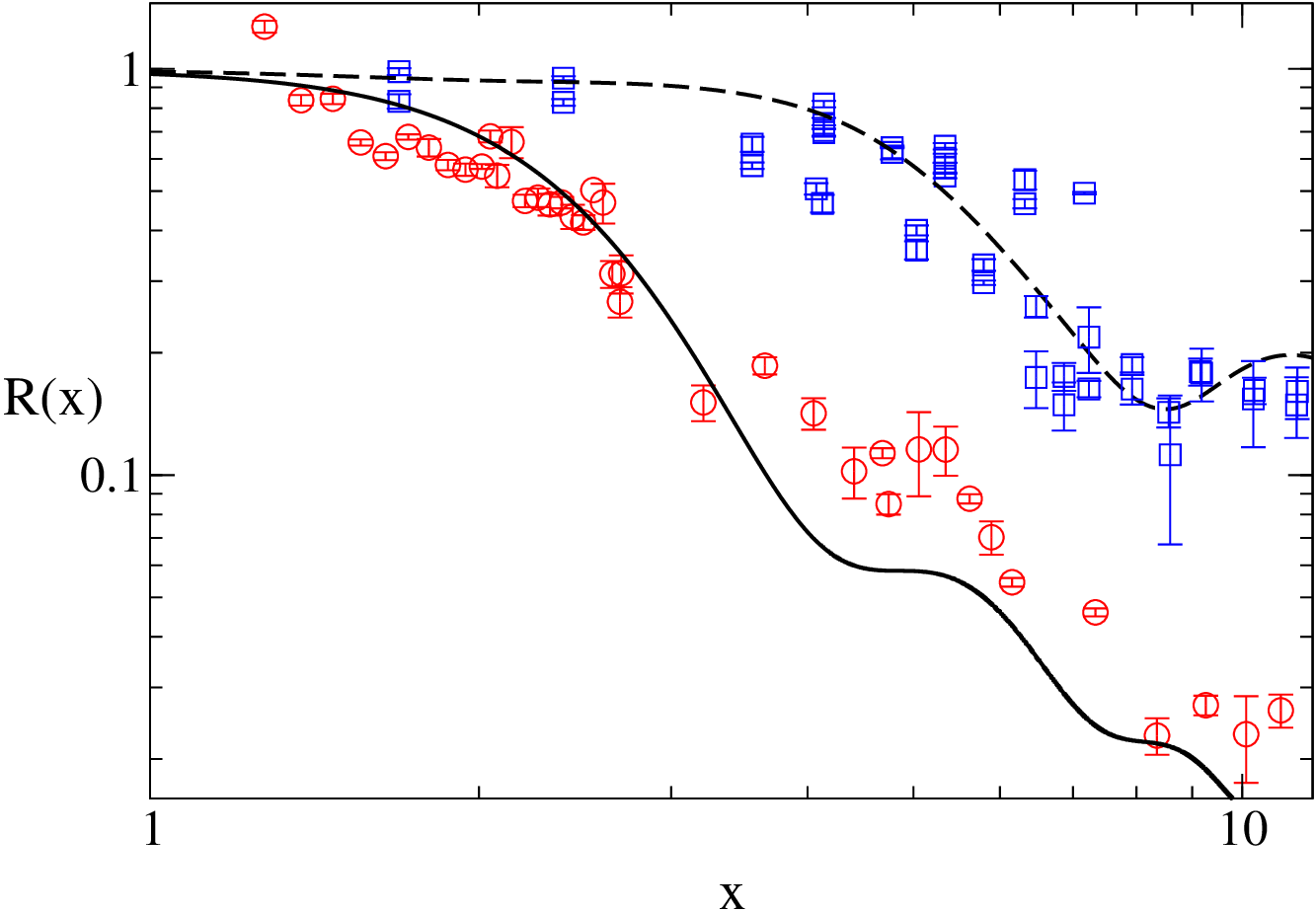} 
\caption{ 
Rescaled experimental mean energies  
near classical resonance (circles),  
and the quantum resonances 
at $\tau=2\pi$ and $4\pi$ (squares).  
In particular, the circles for $x \lesssim 3$  
are rescaled data from the  
inset of Fig. \ref{fig:3} (a). The mean energies have been scaled 
by the theoretical peak height of the  
resonances, i.e., by $k^2t^2/4$ for the classical resonance  
\cite{Sadgrove2005} and $k^2t/4$ for  
the quantum resonance data \cite{Wimberger2003,Wimberger2004}.  
The scaling functions for the classical  
(\ref{eq:scalcl}) and quantum resonances (\ref{eq:scal}) 
are shown as a solid line and a dashed 
line respectively. The narrower  
width of the classical resonance peak is  
immediately apparent. This figure also shows the utility 
of the scaling function in the comparison  
of data which is meaningful in the scaled units  
even for a wide range of the three parameters: here for  
$k\approx 2.5$ and $k\approx 5$ ($\ell=0$) and 
$k\approx 5$ ($\ell=1,2$), $0.033 \leq \epsilon \leq 0.1$ ($\ell=0$)  
and $0.03 \leq \epsilon < 0.3$ ($\ell=1,2$), and $3 \leq t \leq 16$. 
Error bars represent statistical fluctuations over three independent 
experiments. 
}  
\label{fig:4} 
\end{figure} 
 
Lastly, Fig. \ref{fig:4} shows rescaled data from  
experimental measurements for 
various experimental parameters with $\ell=0$  
(circles) and $\ell=1,2$ 
(squares). The data taken in the  
classical case ($\ell=0$) falls on or close to 
the classical scaling function (solid line in  
Fig.~\ref{fig:3}) and that, likewise, the data  
taken for $\ell=1,2$ falls on or near the quantum  
scaling curve (dash--dotted line). The narrower 
nature of the classical resonance peak is emphasised  
by this plot. The dense set of points (circles) 
shown for $x\lesssim 3$ in the classical case come  
from the data shown in the inset of  
in Fig.~\ref{fig:3} (a).  
This data provides a detailed confirmation of the  
classical scaling function's validity for smaller  
values of $x$ than previously observed experimentally \cite{pra2005}.  
Somewhat surprisingly, it is found that the $\delta$--kicked rotor
theory holds even in a regime of $x$ for which the spacing between
kicking pulses is comparable to the width of the pulses themselves
\cite{Sadgrove2005}. The smallest value of the kicking period $\tau$
for which the $\delta$--kicked model remained valid in these experiments was
$\tau = 0.033$ which, for a kicking strength $k\approx 5$ and $t=5$,
corresponds to $x\approx 2$. 
For larger $x$, the data points show more scatter  
because of systematic fluctuations in the initial  
momentum spread and the difficulty in observing the  
peak very close to resonance for a larger 
number of kicks \cite{pra2005}. 
 
\section{Conclusion} 
 
We have demonstrated the effect of averaging over  
a uniform quasi--momentum distribution in two 
different semiclassical limits of the atom--optics kicked rotor. 
For the {\em same} experimental quasi--momentum 
distribution, the true classical limit gives rise to  
{\em ballistic} energy growth whereas 
in the pseudo-classical limit approximating quantum  
resonance only linear growth occurs.  
 
This difference is explained by considering the  
inclusion of the quasi--momentum dependent term  
$\tau \beta$ in the theoretical description. If this  
term approaches zero, which may be accomplished 
either by performing the classical limit $\tau \to 0$  
or starting with a very narrow momentum 
distribution such as that provided by a Bose-Einstein  
condensate \cite{pisa},  
ballistic energy growth is recovered.  
However, for standard atom--optics kicked rotor  
experiments using cold atoms only linear 
energy growth is predicted at quantum resonance  
since the quasi--momentum $\beta$ is  
uniformly distributed in the entire Brillouin zone. 
 
The classical theory of Section \ref{drei} of the near  
resonant dynamics thus unifies the description 
of quantum and classical resonance behaviour  
of the  atom--optics kicked rotor, and is elegantly summarised by two  
classical one-parameter scaling laws 
for the classical and quantum resonance  
peaks. These laws are very useful for a detailed analysis
of experimental results in regimes in which measurements 
are limited by the signal-to-noise ratio.

\ack 
The authors would like to thank Rainer Leonhardt and Scott  
Parkins for helpful discussions. Additionally, we are grateful 
to Andreas Buchleitner and Javier Madro\~nero for their  
hospitality and logistical support at the Max Planck  
Institute for the Physics of Complex Systems in Dresden. 
M.S. was supported by The Tertiary Education  
Commission of New Zealand. S.W. thanks the organisers of the 
International Workshop on ``Aspects of Quantum Chaotic Scattering''
(Dresden, 2005) for providing a stimulating atmosphere 
and partial financial support, as well as the 
Alexander von Humboldt Foundation (Feodor-Lynen Program) for funding.

\end{document}